\def\e{\varepsilon}

 \documentstyle[11pt]{article}
\newcommand{\lsim}{\raisebox{-.025in}{$\stackrel{<}{{\scriptstyle \sim}}$}}

\def\sqr#1#2{{\vcenter{\vbox{\hrule height.#2pt
\hbox{\vrule width.#2pt height#1pt \kern#1pt
\vrule width.#2pt}
\hrule height.#2pt}}}}
\def\square{\mathchoice\sqr34\sqr34\sqr{2.1}3\sqr{1.5}3}

\begin{document}
\thispagestyle{empty}
\begin{center}

\hbox{}
\vspace{0.2in}

{\LARGE \bf A Criterion That Determines Fast Folding of Proteins:\\
A Model Study}

\vskip.4in 

Carlos J. Camacho\footnote{e-mail: ccamacho@lascar.puc.cl}\\
\vspace{0.1in}

{\em Facultad de
F\'\i sica, P. Universidad Cat\'olica de Chile, Casilla 306, Santiago
22, Chile}\\
\vspace{0.1in}

D. Thirumalai\footnote{e-mail: thirum@glue.umd.edu}\\
\vspace{0.1in}
{\em Institute for Physical Science and Technology,
 University of Maryland, College Park, MD 20742}\\

\vspace{0.6in}

{\bf Abstract}
\end{center}
\setlength{\baselineskip}{24pt}

We consider the statistical mechanics of a full set of 
two-dimensional protein-like
heteropolymers, whose
thermodynamics is characterized by the
coil-to-globular ($T_\theta$) and the folding ($T_f$) transition 
temperatures. For our model, the typical time scale for reaching
the unique native conformation is shown to scale as 
$\tau_f\sim F(M)\exp(\sigma/\sigma_0)$,
where $\sigma=1-T_f/T_\theta$, $M$ is the number of residues, and
$F(M)$ scales algebraically with $M$. We argue that $T_f$ scales
linearly with the inverse of entropy of low energy non-native states,
whereas $T_\theta$ is almost
independent of it. As $\sigma\rightarrow 0$, non-productive
intermediates decrease, and the initial rapid collapse of
the protein leads to structures resembling the 
native state. 
Based solely on {\it accessible} information,
$\sigma$ can be used to predict sequences that fold rapidly.

\vspace{0.6in}
\vspace{0.6in}
\pagebreak
\vspace{0.5in}
\setlength{\baselineskip}{24pt}

An apparent puzzle in the protein folding kinetics was raised by Levinthal
\cite{LEV}
in late sixties. He argued that since the number of conformations of even
a moderate sized protein is astronomically large, it is unlikely that a
polypeptide chain can
find the lowest free energy conformation (referred to as the native state)
in biologically relevant time scales. We note that the time
scale in which proteins fold in cells is several (twelve or more) orders of
magnitude longer than microscopic time scales. The belief that
proteins find the global free energy minima in times on the order of
seconds led Levinthal to postulate 
that there must be ``preferred pathways'' that
direct the folding process. Minimal protein models, which capture some
but not all of the features considered to be important in proteins, have in
recent years been used to provide plausible resolutions to this seeming
paradox
\cite{JBPW,PNAS1,PRL,EIS}. These scenarios have many common elements
but differ significantly in detail. The unifying idea that has emerged from
these studies is that in order to quantitatively describe folding kinetics
(at least in these models) one has to contend with complex energy landscapes.

A few years ago we showed using simple two dimensional lattice models that,
in general, quasi random sequences reach their native conformation by a
three stage multipathway kinetics
\cite{PNAS1,PRL,PNAS2}. In the first stage the chain collapses
from a random coil to a compact state. In the second stage the chain
searches among the set of compact structures, in a diffusive reptation like
mechanism, to reach one of minimum energy structures. The final stage
involves an activated transition from one of the minimum energy structures
to the native conformation.
Recently tentative estimates of the time scales
for the various processes in the three stage kinetics have been suggested in
terms of $M$ the number of amino acid residues in a protein and other
experimentally controllable parameters
\cite{PNAS1,DT,CJC}. These estimates support our earlier
assertion that this three stage kinetics provides a resolution to the
Levinthal paradox for single domain proteins which have typically $M$ less
than about two hundred. The reason that folding of these single domain
proteins occurs in time scales on the order of seconds is that the average
free energy barrier in the third stage scales
only as $\sqrt{M}$ \cite{DT} and not as $M$ as had been supposed by others
\cite{MKAS}.  

With the Levinthal paradox resolved by the three stage multipathway
kinetics \cite{PNAS1} for quasirandom sequences, it is natural to address the following
question: For a given value of M is there an intrinsic property of the
sequence that essentially determines the folding time? In this paper we use
a simple model, belonging to the class of HP model proposed by Dill and
collaborators 
\cite{DILL}, to
answer in affirmative the question raised above. 
We had conjectured earlier
\cite{PNAS1}
that sequences that fold rapidly are characterized by having the 
coil-to-globular (collapse) transition temperature, $T_\theta$, and 
the folding temperature, $T_f$, in close proximity. In particular,
the parameter
\begin{equation}
\sigma=(T_\theta - T_f)/T_\theta
\end{equation}
can be used to classify kinetic accessibility of the native conformation.
In this letter, we present quantitative estimates of folding times
in a number of sequences spanning a range of $T_\theta$ and $T_f$
that explicitly verifies our earlier conjecture. This result implies
that kinetic 
accessibility of the native state may in fact
be encoded in the primary sequence of proteins.

The aforementioned prediction can be interpreted in the
context of the refolding of a protein-like structure from an
unfolded conformation ($T>T_\theta$) to a folded native-like structure
($T\,\,\lsim\,\, T_f$). It is reasonable to suggest that $\sigma$ 
probes the key
role played by the ``molten globular'' states in the folding
dynamics. (a) For large $\sigma$, the dynamical process involves
a detailed sampling of transient globular states in a rough free energy
landscape \cite{PNAS1,PRL,DT,CJC}, which naturally
slows down the folding kinetics. On the other hand, (b) for
$\sigma$ small,
the collapse
and folding occur almost simultaneously, with the chain 
collapsing almost directly into a folded structure. 
This was the rationalisation to our earlier prediction \cite{PNAS1}:
the smaller the value of $\sigma$, 
the smaller the value of the folding time scale $\tau_f$! 

The polypeptide chain is modelled as a two letter self-avoiding
walk on a two dimensional square lattice \cite{DILL}. Each bead on 
the chains can be either
hydrophobic (H) or hydrophilic (P). When two non-bonded H beads
are nearest neighbors the interaction is assumed to be $-\e <0$. All
other interactions are zero.
The temperatures $T_\theta$ and $T_f$ are computed {\it exactly}
using series enumeration of the finite-size chains (see, e.g.,
\cite{PNAS1,PRL,DILL,GUT}). The simulation results have been
obtained by 
using a single step Monte Carlo dynamics and a
Metropolis algorithm. All time scales are measured in Monte Carlo
steps (MCS). Allowed moves are such that they mimic
basic features of real chain dynamics, i.e.
preserve chain connectivity and excluded volume interactions
\cite{VER}.
Sequences studied are such that
the ratio of the
number of hydrophobic $M_H$ to
hydrophilic $M_P$ sites is set close to its optimal
folding value of one \cite{PRL}, and they
have a unique ground state assumed to be the native state. Thus,
we restrict
our analysis to (thermodynamically) foldable sequences.
To free our statistical analysis from a bias sampling, we consider
all possible sequences with the aforementioned
properties, and protein sizes equal to
$M=15$ -- $M_H=8$ (214 sequences) and $M=18$ -- $M_H=10$
(1326 sequences).

The time scale $\tau_f$ was measured by fitting and averaging the exponential 
decay ($\exp[-t/\tau_f]$) of the
long-time deviation from equilibrium of several correlation
functions after a temperature quench from a high temperature
unfolded structure to the folding temperature $T_f$. 
For details see Ref. \cite{PNAS1}.
One of the correlation functions used is the overlap function 
$\langle\chi(t)\rangle$ which depends on all distances ${r}_{ij}$ between sites,
with $i$ and $j$ indicating site index along the chain.
This function is a useful probe of the folding kinetics \cite{PNAS1},
and is defined as follows
\begin{equation}
\langle\chi(t)\rangle=1-{1\over M^2-3M+2}\,\,
\langle\sum_{i\neq j,j\pm 1} \delta({r}_{ij}(t)
-{r}_{ij}^N)\rangle\,\,\,.
\end{equation}
It measures structural differences 
between fluctuating conformations and the ground (or 
``native'') state denoted by the superscript $N$.
($\langle\chi\rangle$ varies between 1 for a fully non-native
structure and 0 for the pure native state. Random overlap of two
structures amounts to
$\langle\chi\rangle=0.735$ for $M=15$.)
We denote
non-native states as those structures with, at least, one 
topological feature
different from the native state. 
The folding temperature $T_f$ is defined as the temperature at which
the fluctuations $\Delta\chi=\langle\chi^2\rangle -\langle\chi\rangle^2$
show a peak. 
Below $T_f$ conformations are mostly native,
whereas above $T_f$ they are non-native. 
For a protein model with short range interactions, the standard 
definition for $T_\theta$ is the temperature at which the 
energy fluctuations ($\langle E^2 \rangle-\langle E\rangle^2$) peak,  
these fluctuations are trivially related to 
the specific heat which were used to define $T_\theta$ in Ref. \cite{PNAS1}.
For the cases cited in \cite{PNAS1}, $\sigma\approx
(0.50$, 0.63, $0.088)$, 
whereas $\tau_f\approx (40$, 230, $1) \times 10^5$ MCS,
for models A, B and C, respectively. 
This apparent correlation has  also been recently observed in
three dimensional lattice simulations by Socci and Onuchic 
\cite{NSJO1}. 
These authors 
suggested that reducing the average energetic
drive toward compactness may lead to a smaller difference 
in $T_\theta-T_f$.

The first correlation of interest is between
$\sigma(T_\theta,\,T_f)$ and the folding time scale $\tau_f$.
Figure 1 summarizes this analysis as a function of
the parameter $\sigma$, showing: (a) histograms of $\sigma$
for the space of
sequences mentioned above; (b) the average overlap of the
first exited states with respect to the ground state 
$\langle\chi\rangle_{1st}$; and (c) $\tau_f$ (the equilibration
time scale at $T_f$) for 30 random sequences with $M=15$.
(a) The histograms of $\sigma$ show broad distributions between
$\sigma\simeq 0.2$ and 0.7. 
(b) The average resemblance of the metastable states
($1^{st}$ exited states) with the native structure decreases as
$\sigma$ increases. Indeed, for large $\sigma$ the overlap
is close to that of random structures. Hence, as $\sigma$ increases
metastable states are further apart in configurational space.
(c) Based on the above, it is not surprising to
find that $\tau_f$ varies by almost three orders of magnitude
as a function of $\sigma$. We find the scaling
\begin{equation}
\tau_f\simeq F(M)
\exp[\sigma(T_\theta,T_f)/\sigma_0]\,\,, 
\end{equation}
which we predict to be {\it universal}. 
Although it was not checked here, we expect the prefactor $F(M)\sim 
M^\lambda$
to also be a universal scaling function with $\lambda\simeq 3$, note the
resemblance with Eq. 4 of Ref. \cite{CJC}, and Refs. \cite{PNAS1,DT,DON}. 
The constant 
$\sigma_0\simeq0.11$ is a model dependent parameter. The most important
conclusion of this analysis is that
fast folding sequences are characterized by having small values of $\sigma$.
Some limited experimental verification of this prediction
have recently been shown on fast folding cytochrome {\it c} \cite{SOS}, where
folding and collapse have been found to be almost synchronous.

Since folding times correlates well with $\sigma$ it is 
instructive to find a relationship between $\sigma$ and the
energy spectrum.
Theoretically, we expect
${\cal F}_N(T_f)\approx {\cal F}_{NN}(T_f)$, where
${\cal F}$ denotes free energy and $N$ and $NN$
stand for native and non-native states, respectively. Hence,
one can write the following equation 
\begin{equation}
T_f\approx (\langle U_{NN}\rangle-\langle U_{N}\rangle)/
(\langle S_{NN}\rangle-\langle S_{N}\rangle),
\end{equation}
where $\langle U\rangle$ and $\langle S\rangle$ denote
internal energy and entropy. At $T_f$ it is reasonable to
assume that the leading contribution to the statistical
averages come largely from the low-lying states. Thus, we can estimate
$\langle U_{NN}\rangle-\langle U_{N}\rangle\sim\e$, obtaining
\begin{equation}
\e/k_BT_f\approx \ln(\Omega_{NN}/\Omega_{N})\,,
\end{equation}
where $\Omega$ corresponds to the number of states with
$\Omega_N\sim 2$ (i.e. ground state plus its mirror image), and $k_B$ is
the Boltzmann constant hereafter set to one.
We conclude that $T_f$ must depend linearly on the ``entropy''
of non-native states. We expect (5) to be an excellent estimate for
those sequences with a sparse low energy spectrum.

This prediction
is in very good agreement with Fig. 2, where we show
$\e/T_f$ as a function of the degeneracy of the first $\Omega_1$, second
$\Omega_2$ and third $\Omega_3$ exited states.
For degeneracies larger than 30, $\varepsilon/T_f$ scales
almost linearly with the logarithm of $\Omega_1/2$,
see solid line in Fig. 2. 
The striking agreement between the fit and (5) led us to conclude
that $\Omega_{NN}\simeq \Omega_1$ for large $\Omega_1$.
Although noisier a similar
correlation is observed for higher energy levels as well.
Degeneracies of the energy levels grow exponentially, on an average, by a model
dependent factor of the order of 10 as the energy increases by one.
Clearly, this factor depends
on the physical constraints of the 
connectivity between states. These observations suggest a certain hierarchy
and organization of the energy landscape, with closely related
energy levels. These correlations can be model dependent.

Also shown in Fig. 2 is the small but definite 
dependence of the inverse collapse temperature $\e/T_\theta$
with $\Omega_1$. 
Fast folding sequences have a somewhat
lower $T_\theta$, suggesting that these
sequences have a smaller energetic
drive toward compactness \cite{NSJO1}.
Figure 2 shows the link between the broadening
of $T_\theta-T_f$ and 
``molten
globular'' states.
For clarity, we have also plotted the histograms for $\Omega_1$ whose
shapes are reminiscent of those in Fig. 1.
As the ratio $M_H/M_P$ deviates from unity, the number
of exited states increases dramatically \cite{PRL}, with $\sigma$ and
$\tau_f$ increasing accordingly. Hence, this analysis
presents further evidence regarding the 
role played by 
intermediate states ($\Omega$) in the folding dynamics, and
the natural selection of proteins with an optimum
content of hydrophobic residues \cite{PRL}.

It is noteworthy that the energy gap between the native state
and metastable states in this model is always $\e$. A trivial check of 
correlations between $\e$ and $\tau_f$ shows that there is none.
This, of course, contradicts an earlier suggestion of
\v{S}ali, Shahknovich and Karplus \cite{NAT} who suggested
that the energy gap 
was enough to predict chain ``foldicity'', or ease to fold.
The gap appropriately divided by $T_f$ does seem to show a
correlation with $\tau_f$ \cite{GUT1}. This correlation, however,
deteriorates for fast
folding sequences governed by entropy \cite{PROT}.

Goldstein {\it et. al.} \cite{GOLD} proposed that rapid folding sequences are 
characterized by having a large value of the ratio $T_f/T_g$, where
$T_g$ is an equilibrium transition temperature.
From a practical point of view, however, this prediction 
which appears to be supported by simulations \cite{NSJO},
provided $T_g$ is replaced by a kinetic glass transition temperature,
is not very useful. Indeed,
there is no straightforward technique to measure the glass
transition temperature, other than
to estimate $T_g$ from a detailed knowledge of
the unknown $\tau_f$.

It is quite clear that based on two thermodynamic parameters 
one cannot fully describe the folding dynamics in a
complex energy landscape. The above notwithstanding, one
can establish meaningful {\it statistical} relationships between
equilibrium properties and a given set of dynamical rules.
As long as these rules mimic essential features of the
physical processes, the relationships should shed
some light on the underlying mechanisms.

We have shown that experimentally accessible
information, namely $\sigma =1- T_f/T_\theta$ where
$T_\theta$ is the coil-to-globular and $T_f$ is the folding transition
temperature, could be used to predict and design fast folding sequences of
proteins. For sequences that fold rapidly we predict that folding
times $\tau_f$ should scale as suggested by (3).
This expression embodies the interplay between 
energy frustration and entropic barriers. 
It recovers the slow folding limit 
when the protein size $M$ increases and few
sequences fold fast. 
Our postulates are physically limited by its statistical nature. In particular,
averaging over sequence randomness entails standard deviations in
$\tau_f$ of the order of half a decade.
From a microscopic point of 
view, $\sigma$ appears to be related to entropy of low-lying states.
Minimally
frustrated fast folding
sequences with relatively small $\sigma$ fold in time scales mostly governed
by entropic considerations.

\centerline{$\ast\ast\ast$}

This work was supported in part by a grant from the National Science
Foundation (CHE93-07884) and the Air force office of Scientific
Research (F496209410106).
CJC acknowledges support from FONDECYT No. 3940016 (Chile).

\newpage
\noindent{\Large \bf Figure Captions}
\begin{enumerate}
\item
Statistical analysis of $\sigma$ for non-degenerate sequences
with $M=15$ and $M=18$ (see text).
(a) Bottom, histograms of $\sigma$: dashed line
and solid lines correspond to $M=15$ and $M=18$, respectively.
(b) Middle, average overlap between all first exited states and
the ground state 
$\langle\chi\rangle_{1st}$: $+$ and $\diamond$ symbols correspond
to $M=15$ and $M=18$, respectively. Least square fit shows
a linear dependence of $\langle\chi\rangle_{1st}$ on $\sigma$ (solid line). 
(c) Top, folding time
scale $\tau_f$ ($\square$) as a function of $\sigma$ for 30
random sequences with $M=15$. Error bars are of the order of
symbol size. Least square fit yields
$1.8\times10^5\exp(\sigma/0.11)$ (solid line).
Bars in (b) and (c) correspond to
one standard deviation.

\item
Inverse of folding temperature as a function of the number of
first ($\Omega_1$), second and third exited states. Solid line
corresponds to $\e/T_f=\ln(\Omega_1/2)$. Symbols are as in
Fig. 1b. To better resolve the overlap of symbols, we show
histograms of $\Omega_1$ for $M=15$ (dashed line) and $M=18$
(solid line). Also shown is inverse of collapse temperature $\e/T_\theta$
($\times$) as a function of $\Omega_1$.

\end{enumerate}

\end{document}